\documentclass[%
reprint,
showpacs,
showkeys,
 amsmath,amssymb,
 aps,
pra,
]{revtex4-1}

\usepackage{graphicx}
\usepackage{dcolumn}
\usepackage{bm}
\usepackage{hyperref}

\usepackage{siunitx}
\usepackage{xspace}
\usepackage{multirow}

\graphicspath{{Figures/}} 
\newcommand{\ba}{$^{138}$Ba$^+$\xspace}

\begin{document}

\title{Co-trapping different species in ion traps using multiple radio-frequencies}
\date{\today}

\author{Dimitris Trypogeorgos}
\altaffiliation[Currently at: ]{Joint Quantum Institute, University of Maryland, College Park, Maryland,
20742, USA}
\email[E-mail: ]{dtrypo@umd.edu}
\author{Christopher J. Foot}
\affiliation{Clarendon Laboratory, Department of Physics, University of Oxford,\\Parks Road, Oxford, OX1 3PU, UK}

\begin{abstract}
We consider the stability of systems subjected to periodic parametric driving such that their equations of motion are ordinary differential equations with periodic coefficients and carry out a detailed analysis of important aspects of such systems in the context of the confinement of ions by oscillating electric fields. We show how they can be understood in terms of a pseudopotential approximation and resonances arising from the parametric excitation and investigate the properties of a novel linear Paul trap configuration operating with two radio-frequencies to simultaneously confine two species with extremely different charge-to-mass ratios. The theoretical calculations have been verified by molecular dynamics simulations and normal modes analysis.
\end{abstract}



\maketitle

\section{Introduction}

Paul traps have been incredibly successful at confining charged particles, ranging from atomic ions to macroscopic charged objects~\cite{itano_cooling_1995, willitsch_chemical_2008}. The intrinsic sensitivity of the Paul trap mechanism on the charge-to-mass ratio of trapped particles can be exploited for mass spectrometry~\cite{march_quadrupole_2005, ghosh_ion_1996}. While it is possible to simultaneously confine species with markedly different charge-to-mass ratios in a Paul trap, this results in the more weakly confined species being pushed away from the trap centre~\cite{offenberg_translational_2008}. In this article, we describe a method of operating a Paul trap, with two radio-frequency (RF) driving fields, in which two ions of \textit{extremely different} charge-to-mass ratios experience harmonic pseudopotentials with similar spring constants~\cite{[{A preliminary discussion on two-frequency traps appears in }] dehmelt_economic_1995}. Both species can be tightly confined near the centre of the trap so that their interactions are enhanced. 

Gas-phase chemistry of ions at millikelvin temperatures concerns the study of reaction rates and thermodynamic properties in the regime where quantum effects are important. Previous experiments and numerical simulations have used single frequency Paul traps and were limited to species with similar charge-to-mass ratios. Typically, the lighter species is a singly charged atomic ion, with a suitable laser cooling transition, co-trapped with a small molecule with mass of a few hundred amu~\cite{ostendorf_sympathetic_2006}. Our scheme opens the way to working with much heavier charged particles such as nanodiamonds and graphene~\cite{kane_levitated_2010}.
Paul trapping of large biomolecular ions has also been discussed for DNA, but in an aqueous solution rather than the usual vacuum environment~\cite{zhao_molecular_2008}.  

The article is organised as follows. Section~\ref{sec:paul} summarises the well established theory of operation of the Paul trap in terms of the Mathieu equation and the pseudopotential approximation. 
The conditions for stable trapping are calculated using Floquet theory. We discuss parametric resonance and introduce the concept of a critical line that separates stable from unstable regions. The functional form of the critical line is affected by linear damping. In Section~\ref{sec:twofreq} we turn to the two frequency Hill equation and show how the spring constants of two co-trapped species can be independently adjusted. In Section~\ref{sec:example} we present an example of co-trapping and sympathetically cooling a large charged particle with a handful of atomic ions. We use molecular dynamics and normal modes analysis to simulate the full dynamics of the system. Finally, we conclude in Section~\ref{sec:conc} with a discussion of the experimental relevance of our results and future prospects.

\section{Kinematics of the Paul trap}
\label{sec:paul}

The Paul trap, invented by Wolfgang Paul, confines ions with an oscillating electric quadrupole field~\cite{[{Wolfgang Pauli humorously referred to him as his ``imaginary part'', }] brown_hans_2006, paul_electromagnetic_1990-1}. It has many diverse applications including frequency standards and quantum computing. The development of laser cooling techniques had an important impact on the use of ion traps, and even charged particles without suitable optical transitions can be cooled sympathetically by exchanging energy with atomic ions, amenable to laser cooling techniques. 

In the following we consider a linear Paul trap with alternating voltage $V(t)=V_1\cos(\Omega_1t)$ applied to four parallel rod-shaped electrodes and the end-caps held at a constant voltage $U_0$. The total electric field produced by such an arrangement is
\begin{equation}
\mathbf E(\mathbf x; t) = -V(t)\left(\frac{x\hat x-y\hat y}{R_0^2}\right) + \frac{U_0}{Z_0^2}(x\hat x  + y\hat y - 2z\hat z)
\label{eq:full}
\end{equation}
where $R_0$, $Z_0$ are the characteristic lengths along the radial and axial directions respectively. 

\subsection{Mathieu equation}

\begin{figure}[ht]
\centering
\includegraphics[width=\columnwidth]{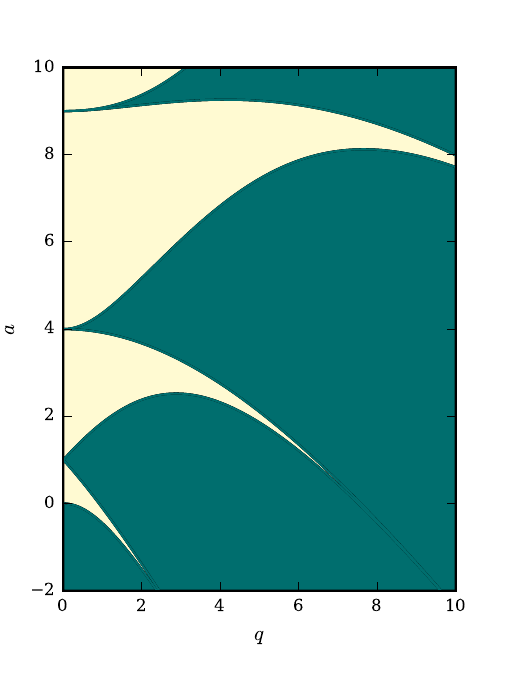}
\caption{The stability diagram of the Mathieu equation. The lightly coloured regions indicate where the system is stable. The diagram corresponds to a single transverse axis of a linear Paul trap. Full radial confinement is possible only at the intersections of stable regions for the remaining axis. The usual operating regime for linear ion traps lies very close to the origin $a=0$ axis~\cite{ghosh_ion_1996}.}
\label{fig:1}
\end{figure}

The equation of motion for a single charged particle in the electric field of Eq.~\ref{eq:full}, including linear damping $\beta$, can be written in the form of a Mathieu equation~\cite{paul_electromagnetic_1990, wineland_experimental_1998}
\begin{equation}
\frac{d^2u_i}{dt_1^2}-\beta_1\frac{du_i}{dt_1}+\left(a_i-2q_i\cos{\left(2t_1\right)}\right)u_i=0
\label{eq:mathieu}
\end{equation}
where $2t_1=\Omega_1 t$ is the effective time, $i$ indexes the three spatial coordinates and $\beta_1 = 2\beta / \Omega_1$. The constants $a_i$ and $q_i$ depend on the ratio of charge $Q$ and mass $M$ of the trapped ion and the amplitudes of the static and oscillating voltage components: 
\begin{equation}
a_x=a_y=-\frac{1}{2}a_z=-\frac{4QU_0}{MZ_0^2\Omega_1^2}
\end{equation}
and similarly for $q$:
\begin{equation}
q_x=-q_y=\frac{2QV_1}{MR_0^2\Omega_1^2}, \quad q_z=0
\label{eq:qq}
\end{equation}

The non-singular solutions of Eq.~\ref{eq:mathieu} are the Mathieu characteristic functions~\cite{arfken_mathematical_2005,ruby_applications_1996,mclachlan_theory_????}. They are the elliptic equivalent of the trigonometric functions and were first discussed by Mathieu in 1868~\cite{mathieu_memoire_1868, whittaker_course_2009} in the context of finding the normal vibration modes of an elliptic membrane. An approximate solution to Eq.~\ref{eq:mathieu} can be found by separating the fast and slow oscillating components of the atomic motion, assuming the amplitude of the fast component is small. This is referred to as the \textit{pseudopotential approximation}. The frequency of the secular motion of the ion is then related to the driving frequency by
\begin{equation}
\omega_i\simeq\frac{\Omega_1}{2}\sqrt{a_i+\frac{1}{2}q_i^2}
\label{eq:freqs}
\end{equation}

In the above we have assumed that the ion is unconditionally stable which is true for $a\ll1$ and $q\le0.9$. This constitutes the first stability region of the Mathieu equation where ion traps normally operate. Fig.~\ref{fig:1} shows the stability diagram for the one-dimensional Mathieu equation which is calculated using Floquet theory as described in Appendix~\ref{sec:floq}. Note that for the remainder of the article we only consider motion and stability along the $y$-axis and drop the $i$ subscript from the Mathieu coefficients.

\subsection{Damping and critical lines}
 
\begin{figure}[ht]
\centering
\includegraphics[width=\columnwidth]{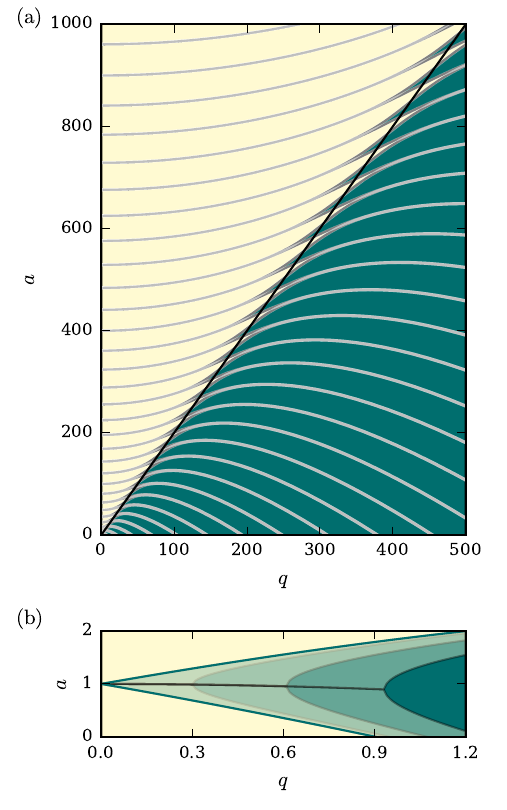}
\caption{(a) The critical line that passes through stable and unstable regions of equal density is $a=2q$ for the undamped Mathieu equation. (b) The first parametric resonance located at $a=1$.  The addition of damping lifts the parametric resonances from the vertical axis enlarging the stability region. The different shaded regions correspond to damping values $\beta_1 = 0.3,\,0.6,\,0.9$. The line intersecting all the contours is where the odd and even solutions of the Mathieu equation becoming degenerate.}
\label{fig:2}
\end{figure}

The behaviour of the Mathieu equation in the high $a$-$q$ regime can be mapped to a subharmonics picture which is strongly related to the multi-frequency operation of the Paul trap. The mathematical properties of the Mathieu equation in this regime are well studied although it is not usually relevant to ion trapping.

As can be seen in Fig.~\ref{fig:2}a, when there is a large number of parametric resonances the fine features of the stability diagram are lost. To differentiate between stable and unstable regions we introduce the concept of a critical line which for the undamped Mathieu equation is simply $a=2q$~\cite{broer_geometrical_1995, broer_large_2013, weinstein_asymptotic_1987}. The critical line is defined as the geometric collection of points that separate the stability diagram in regions of equal stable/unstable density. Since the usual point of operation of the Paul trap is close to the vertical axis such a critical line would overestimate the stability region. A more useful critical line can be formulated by including damping.

The addition of damping changes the functional form of the critical line and enlarges the region where the system is stable as shown in Appendix~\ref{sec:floq}. The parametric resonances are lifted from the $a$ axis towards a finite $q$ as can be seen in Fig.~\ref{fig:2}b. The boundaries of the resonance lie along $a=1\pm q$ for $q\ll 1$ and $\beta_1=0$. The boundaries change with added damping and the critical value for exciting the parametric resonance becomes approximately equal to $\beta_1$. The width of the instability tongues is simply the difference between the odd and even eigenvalues of the Mathieu equation which for $q\ll1$ scales as $q^m (1+\mathcal{O}(q^2))$, where the integer $m$ is the order of the resonance emanating from $a=m^2$ on the vertical axis~\cite{NIST:DLMF, Olver:2010:NHMF}. The asymptotic behaviour of the width close to the $q=0$ line indicates that a small value of damping is sufficient to suppress the parametric resonance especially for higher order resonances. The critical value of the parametric driving that is needed to excite a resonance in the damped system is 
\begin{equation}
q\Big|_{\mathrm{crit.}} = C_m \left(\frac{\omega}{\Omega_1}\right)^2 \left(\frac{\beta}{\omega}\right)^{1/m}
\label{eq:qc}
\end{equation}
as has been shown in~\cite{zhao_parametric_2002, razvi_fractional_1998}. Using the series expansion for the difference between odd and even eigenvalues of the Mathieu function the coefficient $C_m = m^{-2}\left(2^{2m-1} (m!)^2\right)^{1/m}$.

\section{Stability of the two-frequency Paul trap}
\label{sec:twofreq}

\begin{figure}[ht]
\centering
\includegraphics[width=\columnwidth]{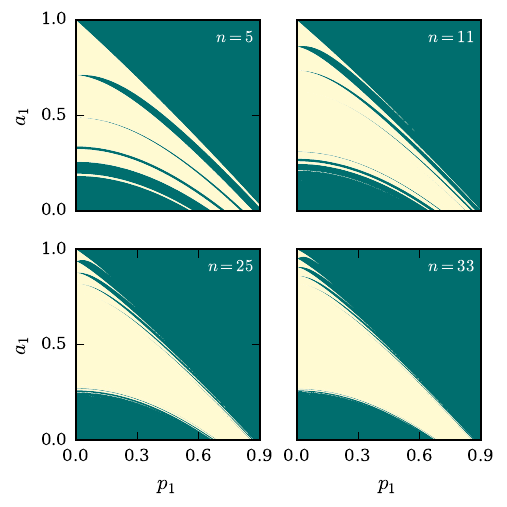}
\caption{The first region of stability of the two-frequency Hill equation on the $a_1$-$p_1$ plane for $n=5,\,11,\,25,\,33$ and $q_1=0.04$. Every panel contains $n-1$ subharmonic resonances which for larger values of $n$ cannot be shown at this resolution, since their width is exponentially small~\cite{konenkov_matrix_2002}.}
\label{fig:aq}
\end{figure}

The equation of motion along the $y$-axis of a trapped ion subjected to an oscillating voltage of the form $V(t)=V_1\cos{\Omega_1t}+V_2\cos{\Omega_2t}$ is
\begin{equation}
\frac{d^2y}{dt_1^2}+\left( a_1 - 2q_1\cos{(2t_1)}-2p_1\cos{(2nt_1)}\right) y=0
\label{eq:twofreq}
\end{equation}
where time has been rescaled so that $\Omega_1t = 2t_1$ and the rest of the parameters are defined as for Eq.~\ref{eq:mathieu} with $q_1 \propto V_1$ and $p_1 \propto V_2$. The subscripts on the parameters  indicate which driving frequency is being used to rescale Eq.~\ref{eq:twofreq}. Here, the second driving frequency is a harmonic of the first $\Omega_2=n\Omega_1$, but fractional relations lead to similar results~\cite{leefer_investigation_2016}. To facilitate the presentation and interpretation of relevant results we consider only moderate values of $n$. The introduction of the subharmonic frequency $\Omega_1$ leads to $n-1$ parametric resonances that slice through the previously stable regions as shown in Fig.~\ref{fig:aq}.

\begin{figure}[ht]
\centering
\includegraphics[width=\columnwidth]{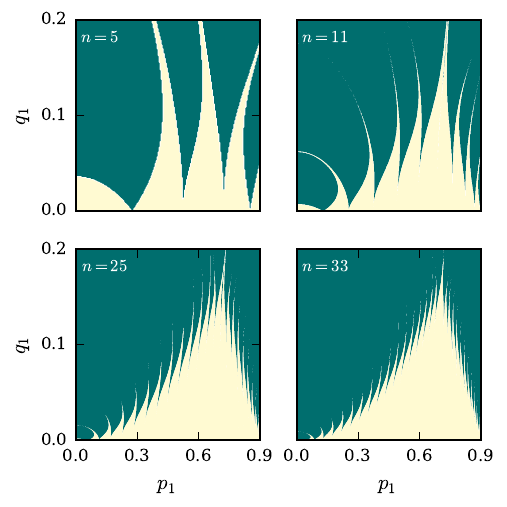}
\caption{The region of stability of the two-frequency Hill equation along $a=0$ on the $q_1$-$p_1$ plane for $n=5,\,11,\,25,\,33$. The system parameters can be chosen so that any parametric resonances are avoided for the top row. This is not the case for larger values of $n$ where the concept of a critical line becomes useful.} 
\label{fig:qp}
\end{figure}

We identify two regimes of operation based on the value of $n$. When $n$ is small the subharmonic resonances are well defined and a stable configuration for the system can be found simply by avoiding them. Damping is less critical in this regime since stable points can be accessed by a judicious choice of system parameters. For higher values of $n$, the system enters a universal regime where the density of the instability tongues cutting through the stable region increases significantly, making it difficult to avoid them. However, the width of the resonances is exponentially small and they can be suppressed by modest amounts of damping as shown by Eq.~\ref{eq:qc}. We then only need to consider the critical line of the system to reason about its stability. A similar picture arises if we take a slice of the stability diagram along the $a_1=0$ line on the $q_1$-$p_1$ plane (Fig.~\ref{fig:qp}).

\subsection{Co-trapping two species}

Having established the criteria for stability we show how to simultaneously trap two species $M_A,\,M_B$ where $M_B>M_A$ with the two ion clouds overlapping. This can be achieved in a single-frequency trap only if the ions have the same $Q^2/M$ since the spring constant is $\kappa=M\omega^2 \propto Q^2/M$. However, the charge-to-mass ratio generally decreases for heavy ions and putting them into a higher charge state might not balance the spring constants because of the discretisation of charge. The required ratio of the dominant spring constants for the clouds to overlap is
\begin{equation}
\frac{\kappa_A}{\kappa_B} = \left(\frac{V_2/\Omega_2}{V_1/\Omega_1}\right)^2 \times \frac{Q_A^2/M_A}{Q_B^2/M_B} \simeq 1
\label{eq:kratio}
\end{equation}

In previous works with single-frequency Paul traps the heavier ions were pushed to the outside of the atomic species cloud since $\kappa_B \ll \kappa_A$~\cite{offenberg_translational_2008, ostendorf_sympathetic_2006}. The two-frequency trap offers a significant improvement since the additional frequency dimension enables for the two species to have the same spring constants $\kappa_A =\kappa_B$, hence similar displacements from the centre of the trapping potential $\vert x_A \vert \approx \vert x_B\vert $ and larger overlap. By adjusting the values of $V_i/\omega_i$ in Eq.~\ref{eq:kratio} we can achieve a precise balance of the spring constants for the two species so that both species congregate within a similar distance from the trap centre since by equipartition of energy, $\kappa\langle x^2\rangle \simeq k_B T$, for ion clouds at temperature $T$.

Let us assume that for $n\gg1$ the system can be thought of as two individual nested Paul traps, since when the two frequencies are sufficiently different there should exist conditions for which one of the oscillating terms can be considered as a minor perturbation acting on the system dominated by the other. The trap can be designed such that $\kappa_B\simeq \kappa_{B,1}$ and $\kappa_A\simeq \kappa_{A,2}$ which indicates that the dominant contribution for the ion $M_B$ comes from the lower frequency $\Omega_1$, and the higher frequency $\Omega_2 \gg \Omega_1$ mostly affects $M_A$. The field at $\Omega_2$ has negligible effect on $M_B$ since ${\kappa_{B,2}}/{\kappa_{B,1}} \ll 1$ and acts to increase trapping in any case. Thus for ions of species B, the two-frequency operation gives a pseudopotential very similar to a standard single-frequency Paul trap. On the other hand, the fact that the variation of the trapping potential corresponding to $\Omega_1$ is slow enough to be adiabatic with respect to $\Omega_2$ implies that $\kappa_A \simeq \kappa_{A,2}$. The quadrupole field at $\Omega_1$ induces parametric resonances at subharmonic frequencies of $\Omega_2$.

Some constraints on the values of $V_i$ and $\Omega_i$ are imposed from the following considerations. The secular oscillation frequency of the light ions is $\omega_A \simeq q_A \Omega_2/(2\sqrt{2})$. The frequency of this mechanical motion sets an upper limit for the lower radio-frequency $\Omega_1$. The first parametric resonance will occur for an excitation at twice the natural frequency of the system and subharmonics thereof. This implies a lower limit of the mass ratio at $M_B/M_A \ge n^2$. This limit can be established more precisely by numerical simulations in specific cases. The radio-frequency $\Omega_1$ is chosen such that $\Omega_1 < \omega_A$ in order to avoid the parametric excitation of species $M_A$ by the radio-frequency field that confines $M_B$. The resulting natural hierarchy of frequencies $\omega_B<\Omega_1<\omega_A<\Omega_2$ leads to two nested Paul traps capable of confining overlapping clouds of two species with very different charge-to-mass ratios.

\section{Example of co-trapping an atomic ion with a biomolecule}
\label{sec:example}

We demonstrate the feasibility of co-trapping atomic \ba ions with $M_A =140$\,amu, $Q_A =1$ and a heavy particle with $M_B=\num{1.4e6}$\,amu, $Q_B=33$, e.g. a nanoparticle or a macromolecule~\cite{benesch_protein_2007}. The choice of this extreme difference in charge-to-mass ratios fully illustrates the potential of this method. We choose $\Omega_2 =2\pi\times\SI{10}{MHz}$ as the main driving frequency. The maximum driving frequency ratio is set by $\Omega_1 <  \Omega_2\times (M_A/ M_B)$. To achieve the same spring constant for both species we choose $n=100$ which makes $\Omega_1=2\pi\times 100$\,kHz. The values of the applied voltages are chosen so that both species nominally experience a single-frequency Paul trap with $q\simeq 0.3$.

\subsection{Rescaling the equations of motion}

Appropriate rescaling of Eq.~\ref{eq:twofreq} is crucial for an intuitive understanding of the system. There are two ways of rescaling the equations of motion for each species, as shown in Table~\ref{tab:mf}. 

\begin{table}[ht]
\caption{Rescaling of the equations of motion for two species trapped in a two-frequency trap. Both species effectively experience a single frequency Paul trap with $q\simeq0.3$ as indicated by the highlighted values in the first and third row.
}
\begin{ruledtabular}
\begin{tabular}{cccc}
\multirow{2}{*}{$M_B,\Omega_1$} & $a_{B,1}$ & $q_{B,1}$      & $p_{B,1}$ \\
						     & -0.0003    & \textbf{0.307} & 10.758 \\

\colrule
\multirow{2}{*}{$M_B,\Omega_2$} & $a_{B,2}$         & $q_{B,2}$          & $p_{B,2}$ \\
						     & \num{-3.35e-8} & \num{2.952e-5} & 0.001 \\

\colrule
\multirow{2}{*}{$M_A,\Omega_2$} & $a_{A,2}$         & $q_{A,2}$ & $p_{A,2}$ \\
						     & \num{-1.03e-5} & 0.009       & \textbf{0.318} \\

\colrule
\multirow{2}{*}{$M_A,\Omega_1$} & $a_{A,1}$ & $q_{A,1}$ & $p_{A,1}$ \\
						     & -0.107       & 94.5         & 3307.4 \\
\end{tabular}
\end{ruledtabular}
\label{tab:mf}
\end{table}

Different physical pictures emerge depending on the choice of effective time $t_1$ or $t_2$ with the coefficients being rescaled with $n^2$, e.g., $q_{A,2} = q_{A,1}/n^2$. Species $M_B$ is mainly confined by the electric field at $\Omega_1$ with $q_{B,1}\simeq0.3$ as shown in the first row of Table~\ref{tab:mf}. Although the coefficient in front of the high frequency term has a larger value it does not significantly contribute to the trapping. This becomes clear if we omit the $\Omega_1$ term and rescale time to $2t_2 = \Omega_2t$. The equation of motion for the species $M_B$ is then the Mathieu equation
\begin{equation}
\frac{d^2y_B}{dt_2^2}+\left( a_{B,2}-2p_{B,2}\cos{(2t_2)}\right) y_B=0
\end{equation}
with $p_{B,2}=0.001$, which provides negligible trapping and the spring constant arising from this equation is $\kappa_{B,2}\ll \kappa_{B,1}$ (Table~\ref{tab:mf} second row). The stability region for $M_B$ is thus only slightly perturbed by the presence of the high frequency term and the trapping is predominantly due to the $\Omega_1$ term.

In a similar manner, $M_A$ is mainly trapped due to $\Omega_2$ since $q_{A,2}\ll p_{A,2}$ (Table~\ref{tab:mf} third row). The effect of $\Omega_1$ on species $A$ however, is more difficult to assess using Eq.~\ref{eq:twofreq} since $\Omega_1$ now acts as a subharmonic term that can excite parametric resonances. The subharmonics picture is complemented by the rescaling corresponding to the fourth row of Table~\ref{tab:mf}. The large values of the coefficients are potentially misleading but, by treating the fast oscillating term as an effective harmonic pseudopotential term that arises from a static quadrupole, we can map Eq.~\ref{eq:twofreq} to the Mathieu equation
\begin{equation}
\frac{d^2y_A}{dt_1^2}+\left(a_{\mathrm{eff}}-2q_{A,1}\cos{(2t_1)}\right) y_A=0
\end{equation}
where $a_{\mathrm{eff}}$ indicates the effective static potential term. This points to a mapping of the system to the high $a$-$q$ regime where parametric resonances are self-induced. The width of these resonances introduced by the subharmonics decreases rapidly with increasing order number $m$ and thus the higher order resonances are readily suppressed by weak damping.

\begin{figure*}[ht]
\centering
\includegraphics[]{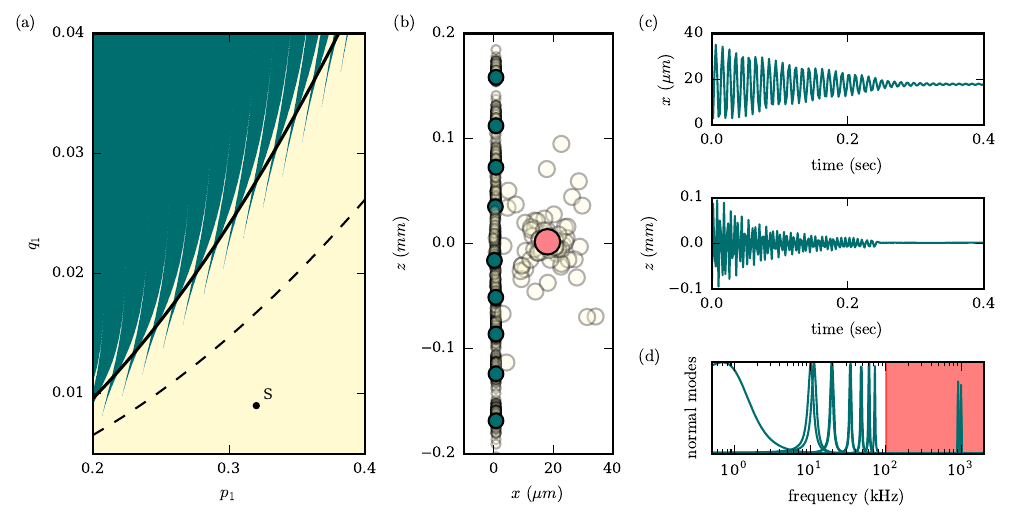}
\caption{(a) Part of the stability region for the \ba ions. The point $S=(0.318, 0.009)$ corresponds to the parameters used to give stable confinement. The tongues of instability extend from the cusps down to the horizontal axis, however, they are not visible due to the finite resolution of the plot even in the absence of any applied damping. The critical line (solid black) is the best fit line $q_1 = -0.002 + 0.29p_1^2$ for this region. The dashed line is the critical drive needed to excite a resonance as calculated by Eq.~\ref{eq:qc2}. (b) Example arrangement of the ions when the system is crystallised. The heavy ion (red) is displaced from the chain of the lighter ones due to $E_\perp$. (c) Characteristic trajectories of the heavy ion. Both axial and radial motion are damped at similar timescales of a few hundred ms. (d) Normal modes of the system. The mid-frequency normal modes are not allowed to cross into the region highlighted in red to avoid parametric excitation of $\Omega_1$.}
\label{fig:100}
\end{figure*}

\subsection{Stability diagram}

Fig.~\ref{fig:100}a shows a portion of the stability diagram of the system for $n=100$. The critical line can be calculated analytically to be $q_1 = 0.27 p_1^2$ for $0< p_1<0.7$. Shown in Fig.~\ref{fig:100}a is the best fit line $q_1 = -0.002 + 0.29p_1^2$. Numerical calculations for $0.7< p_1<0.9$ show that the critical line has a maximum at $ p_1=0.7$ and can be approximated by $q_1=0.57 (0.9 - p_1)$. The stability region extends up to $ p_1=0.9$ as in the Mathieu equation. Parametric excitation of $M_A$ by $\Omega_1$ produces tongues of instability which are too fine for the numeric calculations to capture. Since the light ions experience a pseudopotential with secular frequency $\omega \approx p_{A,2} \Omega_2 / (2 \sqrt2)$ we can use Eq.~\ref{eq:qc} to put an upper limit to the amplitude of the $\Omega_1$ term 
\begin{equation}
q_{A,2}\Big|_{\mathrm{crit.}} = \frac{2p_{A,2}^2}{e^2} \left( \frac{\pi\beta}{\Omega_1} \right)^{1/m}
\label{eq:qc2}
\end{equation}
where the order of the resonance is $m = p_{A,2} \Omega_2 /(\sqrt2 \Omega_1)$ and we have used Stirling's approximation for $m!$. For $n=\Omega_2/\Omega_1=100$ and $p_{A,2}=0.32$ the parametric resonance happens at $m=23$. For this high order resonance even light damping $\beta/\Omega_1 = 10^{-6}$ leads to a critical value $q_{A,2} = 0.016$ which is comfortably above our chosen
operating value $q_{A,2} = 0.01$ (see Fig.~\ref{fig:100}a). For even larger driving frequency ratios, the lower frequency can be considered as DC compared to the secular motion of the lighter species leading to the stability criterion $q_{A,2}<0.5p_{A,2}^2$.

\subsection{Numerical results}

We have carried out extensive numerical simulations using the open-source software package LAMMPS~\cite{plimpton_fast_1995} and verified that the ions are confined in all three dimensions by a quadrupole field oscillating with two frequencies. Using the previous single-ion analysis as a guide for the stability region of the system we simulated the full dynamics of the equations of motion
\begin{equation}
M_j \ddot{\mathbf{x}}_j = \left(E(\mathbf x_j; t) + E_\perp\right)Q_j + \sum_{\substack{i,j=1\\i\neq j}}^N\frac{1}{8\pi\epsilon_0}\frac{Q_iQ_j}{(\mathbf x_j-\mathbf x_i)^2}
\end{equation}
for ion $j$, where $E(\mathbf x_j; t)$ is the electric field from Eq.~\ref{eq:full} with $V(t)=V_1\cos{\Omega_1t}+V_2\cos{\Omega_2t}$,  $E_\perp$ is an optional transverse electric field and $\epsilon_0$ is the permittivity of vacuum. The full-range Coulomb interaction is included. 

For our choice of parameters, the radial and axial oscillation frequencies for $M_B$ are $\SI{11.4}{kHz}$ and $\SI{0.7}{kHz}$ respectively. For $M_A$, the radial frequency is $\omega_A=2\pi \times \SI{1}{MHz}$ and $\omega_{A,z}=\omega_B$ so as to promote resonant energy transfer between the two species. This frequency matching is easily achieved when the number of ions and hence the number of normal modes of the system becomes larger.

The radial electric field $E_\perp$ displaces the ions from the RF-null nodal line and they experience larger micromotion which increases the overall temperature of the system (Fig.~\ref{fig:100}b). However, the crystallisation temperature of the Coulomb crystal is higher than typical atomic systems owing to the much larger Coulomb interactions. The transition temperature can be estimated using the Lindemann criterion, by equating the average amplitude of the thermal vibrations of the ions to their mean separation which is larger by a factor of $Q_B=33$ over single species atomic crystals. Temperatures of a few mK are already low enough for the system to be crystallised. 

We have simulated various configurations involving up to 10 Doppler cooled \ba ions and a single ion of species B. We observed that the damping rate depends strongly on the size and degree of asymmetry of the Coulomb crystal. The axial asymmetry of the crystal was varied by changing the initial position of the heavy ion. More asymmetric crystals tend to cool down faster with cooling times in the range of 100\,ms to 1\,s, inversely proportional to the number of ions. The field $E_\perp$ can be used to improve the coupling of the translational motion of $M_B$ to the normal modes of the light ion chain. Due to the very different secular frequencies of the two species, the out-of-plane motion of $M_B$ is cooled less efficiently and this limits the overall cooling rate. The optimal value for $E_\perp$ can be calculated within the pseudopotential approximation to be in the intermediate regime where a linear chain of $N$ ions transitions to an effective $N-1$ chain with the heavy ion displaced off axis. However, RF micromotion heating limits the maximum value of $E_\perp$ to much lower values.

Fig.~\ref{fig:100}c shows that $1/e$ cooling times of about 200\,ms can be achieved for a crystal of 10 ions. Additional $M_A$ ions position themselves approximately along the z-axis and start forming a chain since $\omega_A \gg \omega_{A,z}$. The axial modes of such a  linear Coulomb crystal have higher frequency than the centre-of-mass mode whereas the opposite is true for higher-order radial modes (see Appendix~\ref{sec:nm}). Adding more ions tends to close the gap in the frequency spectrum of normal modes close to $\Omega_1$ (see Fig.~\ref{fig:100}d). This limits the number of ions of species A that can be accommodated in the same potential well, as an overlap of normal modes with $\Omega_1$ could lead to resonant heating. Larger systems can still be stable by operating at smaller values of $n$ or increasing the damping for all species, e.g. with buffer gas cooling.

Our numerical simulations show how a typical megadalton charged particle can be sympathetically cooled by atomic \ba ions. The presence of the non-fluorescing, dark ion can be deduced from its large effect on the positions of the observable atomic ions. Atomic ions scatter laser light so that individual fluorescing ions can be observed, thus permitting indirect detection of dark ions as holes in the cloud. Pre-cooling can be achieved by means of a conventional single-frequency operation of a linear Paul trap before turning the second frequency on~\cite{ostendorf_sympathetic_2006}. However, with a single frequency the trapping of species B is very weak so these ions might be destabilised by either repulsion from the atomic ions or the radial component of the axial DC field.

\section{Conclusions}
\label{sec:conc}

We presented calculations on the two-frequency operation of an ion trap carried out within the mathematical framework of Floquet theory. Using this as a guide for the stability of the system, we demonstrated the feasibility of  confining different species of ions with the same effective spring constant using molecular dynamics simulations. We specifically chose an extremely different charge-to-mass ratio to demonstrate the usefulness of the two-frequency operation of the trap. Pushing the charge-to-mass ratio even further is possible by applying more than two frequencies. However sympathetic cooling is likely to be more difficult to achieve, as the frequencies of the normal modes become further apart. Our method also works for lower charge-to-mass ratios but it will be more susceptible to normal mode spectral crowding.

Many interesting possibilities arise by being able to extend laser techniques for sympathetic cooling of both the translational and rotational degrees of freedom of large objects that are not amenable to laser cooling likes viruses, molecular motors and dust particles~\cite{benesch_protein_2007, kahra_molecular_2012, staanum_rotational_2010}. Cooling mesoscopic objects like nanodiamonds to their quantum ground state and adapting the sophisticated techniques developed for quantum information processing with trapped ions will allow for investigating entanglement and decoherence dynamics. Reversing the role of the ions, antimatter can be trapped and efficient cooled as has been shown previously only in Penning traps~\cite{jelenkovic_sympathetically_2003}. Our work inspired a re-examination of these ideas for trapping antihydrogen~\cite{leefer_investigation_2016}. 

\begin{acknowledgments}
The authors would like to acknowledge the use of the Oxford Supercomputing Centre in carrying out this work. We acknowledge funding from the Bodossaki Foundation and St. Peter's College (DT) and partial support from the EPSRC. The research leading to these results is supported by EU through the Collaborative Project QuProCS (Grant Agreement No. 641277). We thank Elliot Bentine for useful conversations and reading the manuscript.
\end{acknowledgments}

\appendix

\section{Differential equations with periodic coefficients}
\label{sec:floq}

Ordinary differential equations (ODEs) with periodic coefficients that contain an arbitrary number of frequencies in their Fourier spectrum are ubiquitous in physics. They describe the temporal behaviour of driven systems or the spatial character of the wavefunction in Hamiltonians of crystalline structures, e.g. driven atomic systems, Bloch wavefunctions and mechanical vibrations~\cite{foot_atomic_2004, ashcroft_solid_1976, shirley_solution_1965}. Linear second order homogeneous ODEs with periodic coefficients have the form
\begin{equation}
\frac{d^2u}{dt^2}+B(t)\frac{du}{dt}+S(t)u=0
\label{eq:pode}
\end{equation}
with $B(t+T)=B(t)$ and $S(t+T)=S(t)$ where $T$ is the period of the system. Since any function can be expressed in terms of even and odd functions without loss of generality we choose $S(-t)=S(t)$ and rewrite the $S(t)$ term in Eq.~\ref{eq:pode} equation through its cosine Fourier representation as
\begin{equation}
\frac{d^2u}{dt^2}+\left(c_0+2\sum_{n=1}^\infty c_n\cos{\left(2nt\right)}\right)u=0
\label{eq:cosf}
\end{equation}
where $B(t)=0$. Calculating the characteristic exponents of Eq.~(\ref{eq:cosf}) allows us to map the stability diagram for the phase space spanned by the parameters $c_n$. 
The Mathieu equation is a particular case of Eq.~\ref{eq:cosf} with only the constant and first oscillatory term being non-zero.

\subsection{Floquet theory}

The Floquet formalism can be used to solve ODEs with periodic coefficients~\cite{jordan_nonlinear_2007, konenkov_matrix_2002}. We recast Eq.~\ref{eq:pode} as a system of first order equations
\begin{equation}
\frac{d}{dt}\mathbf X(t)=
\begin{pmatrix}
0 & 1 \\
-S(t) & -B(t) \\
\end{pmatrix}
\mathbf X(t)=0
\end{equation}
where $\mathbf X(t)=(x(t),\dot x(t))^T$. The stability of the system can be analysed by looking at the value of the propagation matrix $M = (\mathbf X_1(T), \mathbf X_2(T))$ after time equal to a period $T$ has transpired. At $t=0$ the propagation matrix equals the identity $M=I$ so that its Wronskian is zero and $\mathbf X_1$, $\mathbf X_2$ are fundamental solutions of the system. The characteristic equation $\det\{M-\lambda I\}$ of the system has solutions $\lambda_{1,2}$ given by its characteristic polynomial
\begin{equation}
\lambda_{1,2}=\frac{1}{2}\left(\mathrm{tr}\{M\}\pm(\mathrm{tr}\{M\}^2-4\det\{M\})^{1/2}\right)
\label{eq:eigenv}
\end{equation}
and the determinant of M is
\begin{equation}
\det\{M\}=\lambda_1\lambda_2=\exp \left(\int_0^T-B(\tau)\mathrm d \tau\right)
\end{equation}

Let us consider the case where $B(t)=0$ so that $\det\{M\}=1$. This is a general property of symplectic matrices that is directly related to the Liouville theorem and expresses the preservation of phase space for a dynamic system. For the system to be stable its eigenvalues must be inside the unit circle in the complex plane $\max\{|\lambda_1|,|\lambda_2|\}\leq1$, or equivalently $|\mathrm{tr}\{M\}|\leq2$.
The trace-determinant plane fully characterises the stability of the system. When $\vert \mathrm{tr}\{M\}\vert > 2$ there are two real eigenvalues. Their product is $\lambda_1\lambda_2=1$ hence either $\lambda_1>1$ or $\lambda_2>1$, giving an unbounded, exponentially diverging solution. On the other hand, when $\vert \mathrm{tr}\{M\}\vert \le 2$ there are two complex eigenvalues, with $\vert \lambda_1\vert =\vert \lambda_2 \vert=1$ and  $ \lambda_1= \lambda_2^*$. They can be written in the form $\lambda = e^{\pm \mathrm i \theta}$ so that $\mathrm{tr}\{M\} = 2 \cos\theta$. These correspond to stable, bounded solutions. In effect, knowledge of the sign of the discriminant $\Delta=\mathrm{tr}\{M\}^2 -4$ is enough to determine the behaviour of the system. The stability points of the undamped Mathieu equation lie along the $\det\{M\}=1$ line and, if stable, are bounded by the parabola $\mathrm{tr}\{M\}^2=4\det\{M\}$. Along the transition curves at the boundary of the above regions $\mathrm{tr}\{M\}= \pm 2$ and the discriminant is zero. The characteristic equation has a double root and degenerate eigenvalues $\lambda_1=\lambda_2=\pm 1$ corresponding to the system oscillating with period $T$ or $2T$ respectively.

\subsection{Damping}

For constant damping $B(t)=\beta>0$, the stability condition becomes  $\vert \mathrm{tr}\{M'\}\vert \leq 1+e^{-\beta T}$ where we have written the propagation matrix as $M'$ to differentiate it from the undamped system. At first sight this equation seems counterintuitive since it appears to reduce the limit on the magnitude of $\mathrm{tr}\{M'\}$. However, the stability region is actually enlarged since $\det\{M'\}=e^{-\beta T}$ rather than 1 as in the undamped case. The product of the eigenvalues in this case is bounded by a circle of radius $e^{-\beta T}$ in the complex plane. The point where one of the eigenvalues becomes greater than 1 is $e^{\beta T}+1$ as can be seen by direct substitution to Eq.~\ref{eq:eigenv}. 

To compare the two cases we can assume that $\lambda$ is the eigenvalue of a matrix of the same form as for the undamped equations $M$, but with $\omega^2\to\omega^2-(\beta)^2$. This leads to the stability condition $\max\{|\lambda_1|,|\lambda_2|\}\leq e^{\beta T}$. The damping term factors out when taking the trace of the propagation matrix, so that $M'=e^{-\beta T} M$. Although the damped system can be investigated directly its behaviour is determined straightforwardly from the corresponding undamped system~\cite{nasse_influence_2001, hasegawa_dynamics_1995}.

\section{Normal modes of motion of damped, stiff systems}
\label{sec:nm}

A system is defined as stiff when at least one of the parameters describing it can take extremely different values that lead to rapid variations in the solution. This is the case when calculating the normal modes of motion of a two-species ion chain where the mass of one species is much larger than the other. Moreover, the addition of damping leads to a matrix equation that is not an eigenvalue equation. Here we show how the matrix equations can be recast to an eigenvalue equation which can be solved using efficient numerical methods.

First we define the pseudopotential for a chain of $N$ ions in a linear Paul trap including the Coulomb interaction
\begin{eqnarray}
V(\mathbf x)&=&\sum_{j=1}^N\frac{1}{2}M_j\omega_{r,j}^2(x_j^2+y_j^2) + Q_jE_\perp x_j + \nonumber \\
&+&\sum_{j=1}^N\frac{1}{2}M_j\omega_{z,j}^2z_j^2+ 
\sum_{\substack{i,j=1\\i\neq j}}^N\frac{1}{8\pi\epsilon_0}\frac{Q_iQ_j}{|\mathbf x_j-\mathbf x_i|}
\end{eqnarray}
with the trap frequencies defined as in Eq.~$\ref{eq:freqs}$. We can find the equilibrium positions by minimising the above potential~\cite{james_quantum_1998}. The field $E_\perp$ not only displaces the atoms along the radial plane but also alters the coupling between the modes. The system is effectively described as a system of coupled oscillators with the restoring forces for each particle arising through the competition between the trapping potential and the Coulomb repulsion of the ions. The equations of motion can be written in matrix form and assuming oscillatory solutions we can calculate the normal modes from the following determinant
\begin{equation}
\det{\{M \omega^2 + G\omega+K\}}=0
\end{equation}
where $M$ is the mass matrix, $G$ is the damping matrix and $K$ is the Hessian of the system. We refer to $G$ as classical damping matrix if $P^T G P$ is diagonal, where  $P$ is matrix of the eigenvectors of the Hessian. The equations of motion are then uncoupled and the damping matrix can be factored into the Hessian and mass matrices.

To treat more generalised (non-classical) damping we follow~\cite{okelly_normal_1961}, whereby the problem of solving the determinant equation is transformed into an eigenvalue problem in the $2N$ space. The normal modes correspond to the eigenvalues of the determinant of the system given by the eigenvalue equation
\begin{equation}
(I\omega^{-1}+U)\mathbf Z=0
\end{equation}
where  $\mathbf Z = \exp(-\omega t) \mathbf X$, $\mathbf X=(\{\dot{\mathbf x}\},\{\mathbf x\})^T$ with dimensions $2N\times 1$ and $I$ is the identity matrix. The block matrix $U$ takes the form
\begin{equation}
U=
\begin{pmatrix}
0 & I \\
-K^{-1} M & -K^{-1} G
\end{pmatrix}
\end{equation}

The eigenvalues of the system correspond to velocity-position pairs. The mode frequencies and characteristic damping times correspond to the inverse of the imaginary and real parts of $\mathbf x$ respectively.

\bibliography{2PaulMethods_Use.bib}

\end{document}